\documentclass[aoas,seceqn,nameyear,dvips]{arximspdf}
\usepackage{dcolumn}
\usepackage{graphicx}

\doi{10.1214/10-AOAS335}
\volume{4}
\issue{3}
\pubyear{2010}
\firstpage{1403}
\lastpage{1429}

\makeatletter
\newcolumntype{d}[1]{D{.}{.}{#1}}

\renewcommand{\citep}[1]{\citet{#1}}
\makeatother

\begin{document}
\begin{frontmatter}

\title{Modeling large scale species abundance with latent spatial processes\protect\thanksref{T1}}
\runtitle{Spatial modeling for species abundance}

\begin{aug}
\author[A]{\fnms{Avishek} \snm{Chakraborty}\corref{}\ead[label=e1]{ac103@stat.duke.edu}},
\author[A]{\fnms{Alan E.} \snm{Gelfand}\ead[label=e2]{alan@stat.duke.edu}},
\author[B]{\fnms{Adam M.} \snm{Wilson}\ead[label=e3]{adam.wilson@uconn.edu}},
\author[C]{\fnms{Andrew M.} \snm{Latimer}\ead[label=e4]{amlatimer@ucdavis.edu}}
\and
\author[B]{\fnms{John A.} \snm{Silander, Jr.}\ead[label=e5]{john.silander@uconn.edu}}

\thankstext{T1}{Supported in part by NSF DEB Grants 056320 and  0516198.}
\runauthor{A. Chakraborty et al.}

\affiliation{Duke University,
Duke University,
University of Connecticut,
University of California, Davis and University of Connecticut}
\address[A]{A. Chakraborty\\
A. E. Gelfand\\
Department of Statistical Science\\
Duke University\\
Durham, North Carolina 27708\\
USA\\
\printead{e1}\\
\phantom{E-mail:} \printead*{e2}} 
\address[B]{A. M. Wilson\\
J. A. Silander, Jr.\\
Department of Ecology and\\
\quad Evolutionary Biology\\
University of Connecticut\\
Storrs, Connecticut 06269\\
USA \\
\printead{e3}\\
\phantom{E-mail:} \printead*{e5}}
\address[C]{A. M. Latimer\\
Department of Plant Sciences\\
University of California\\
Davis, California 95616\\
USA\\
\printead{e4}}
\end{aug}

\received{\smonth{9} \syear{2009}}
\revised{\smonth{2} \syear{2010}}

%
\begin{abstract}
Modeling species abundance patterns using local environmental
features is an important, current problem in ecology. The Cape
Floristic Region (CFR)
in South Africa is a global hot spot of diversity and endemism, and
provides a rich
class of species abundance data for such modeling. Here, we propose a
multi-stage
Bayesian hierarchical model for explaining species abundance over this region.
Our model is specified at areal level, where the CFR is divided into
roughly $37{,}000$
one minute grid cells; species abundance is observed at some locations
within some cells.
The abundance values are ordinally categorized. Environmental and
soil-type factors,
likely to influence the abundance pattern, are included in the model.
We formulate the empirical abundance pattern as a degraded version of
the potential
pattern, with the degradation effect accomplished in two stages. First,
we adjust
for land use transformation and then we adjust for measurement error, hence
misclassification error, to yield the observed abundance classifications.
An important point in this analysis is that only $28\%$ of the grid
cells have been
sampled and that, for sampled grid cells, the number of sampled
locations ranges
from one to more than one hundred. Still, we are able to develop
potential and
transformed abundance surfaces over the entire region.

In the hierarchical framework, categorical abundance classifications
are induced by
continuous latent surfaces. The degradation model above is built on the
latent scale.
On this scale, an areal level spatial regression model was used for modeling
the dependence of species abundance on the environmental factors.
To capture anticipated similarity in abundance pattern among
neighboring regions,
spatial random effects with a conditionally autoregressive prior (CAR)
were specified. Model fitting is through familiar Markov chain Monte
Carlo methods.
While models with CAR priors are usually efficiently fitted, even with
large data
sets, with our modeling and the large number of cells, run times became
very long.
So a novel parallelized computing strategy was developed to expedite fitting.
The model was run for six different species. With categorical data,
display of the resultant abundance patterns is a challenge and we offer
several different views. The patterns are of importance on their own,
comparatively across the region and across species, with implications
for species competition and, more generally, for planning and conservation.
\end{abstract}

%

\end{frontmatter}

\section{Introduction}\label{sec1}

 Ecologists increasingly use species distribution models to
address theoretical and practical issues, including predicting the
response of species to climate change [\citet{Midgley07}, \citet{Fitzpatrick08},
\citet{Loarie08}], designing and managing conservation areas
[\citet{Pressey07}], and finding additional populations of known species or
closely related sibling species [\citet{Raxworthy03}, \citet{Guisan06}]. In all
these applications the core problem is to use information about where a
species occurs and about relevant environmental factors to predict how
likely the species is to be present or absent in unsampled locations.

 The literature on species distribution modeling covers many
applications; there are useful review papers that organize and compare
model approaches [\citet{Guisan00}, \citet{Guisan05}, \citet{Elith06}, \citet{Graham06}, \citet{Wisz08}].
Most species distribution models ignore spatial pattern and thus are
based implicitly on two assumptions: (1) environmental factors are the
primary determinants of species distributions and (2) species have
reached or nearly reached equilibrium with these factors [\citet{Schwartz06}, \citet{Beale07}]. These assumptions underlie the currently
dominant species distribution modeling approaches---generalized linear
and additive models (GLM and GAM), species envelope models such as
BIOCLIM [\citet{Busby91}], and the maximum entropy-based approach MAXENT
[\citet{Phillips08}].
The statistics literature covers GLM and GAM models extensively. The
latter tends to fit data better than the former since they employ
additional parameters
but lose simplicity in interpretation and risk overfitting and poor
out-of-sample prediction.
Climate envelope models and the now increasingly-used maximum entropy
methods are algorithmic and not of direct interest here.

 In addition to the fundamental ecological issues mentioned
above, complication arises in various forms in modeling abundance from
imperfect survey data such as observer error [\citet{Royle07}, \citet{Cressie09}], variable sampling intensity, gaps in sampling, and spatial
misalignment of distributional and environmental data [\citet{Gelfand05a}]. First, since a region is almost never exhaustively
sampled, individuals not exposed to sampling will be missed. Second, it
may be that potentially present individuals are undetected [\citet{Royle07}] and, possibly vice versa, for example, a false positive
misclassification error with regard to species detection [\citet{Royle06}]. A third complication is that ecologists and field workers
are biased against absences; they tend to sample where species are, not
where they aren't. Such \textit{preferential sampling} and its impact on
inference is discussed in \citet{Diggle10}. Further complications arise
with animals due to their mobility. Previous work has developed spatial
hierarchical models that accommodate some of these difficulties,
fitting these models to presence/absence data in a
Bayesian framework [\citet{Hooten03}, \citeauthor{Gelfand05a} (\citeyear{Gelfand05a,Gelfand05b}), \citet{Latimer06}].

 The species distribution modeling discussed above is either
in the presence/absence or presence-only data settings; there is
relatively little work on spatial abundance patterns, despite their
theoretical and practical importance [\citet{Kunin00}, \citet{Gaston03}]. Our
primary contribution here is to develop a hierarchical modeling
approach for ordinal categorical abundance data, explained by the
suitability of the environment, the effect of land use/land
transformation, and potential misclassification error.
Ordinal classifications are often the case in ecological abundance
data, especially for plants [\citet{Sutherland06}, \citet{Ibanez09}]. From a
stochastic modeling perspective, categorical data can be viewed as the
outcome of a multinomial model, with the cell probabilities dependent
on background features. Within a Bayesian framework such modeling is
often implemented using data augmentation [\citet{Albert93}],
introducing a latent hierarchical level. There, the ordered
classification is viewed as a clipped version of a single \textit{latent}
continuous response, introducing cut points. See also \citet{Oliveira00} and \citet{Higgs09}.

 At the latent level, suitability of the environment can be
modeled through regression. Availability in terms of land use degrades
suitability. That is, an important feature of our modeling, from an
ecological point of view, is that it deals with transformation of the
study area by human intervention. In much of the region, the
``natural'' state of areas has been altered to an agricultural or urban
state, or the vegetation has been densely colonized by alien invasive
plant species. So, we cannot treat the entire region as equally \textit
{available} to the plant species we are modeling. We must introduce a
contrast between the current abundance of species (their \textit
{transformed} or \textit{adjusted} abundance) and their potential
distributions in the absence of land use change (\textit{potential}
abundance). These notions are formally defined at the areal unit level
in Section~\ref{sec3}. A further degradation enabling the possibility
of misclassification and/or observer error in the data collection
procedure can be accounted for as \textit{measurement error} in the
latent surface. There is a substantial literature on measurement error
modeling for continuous observations, for example,
\citet{Fuller87}, \citet{Stefanski87}, and \citet{Mallick95}. In our modeling we impose a
hard constraint: with no potential presence (i.e., an unsuitable
environment), we can observe only zero abundance. We enforce this
constraint on the latent scale. With cut points, modeled as random, we
provide an explanatory model for the observed categorical abundance
data. Furthermore, we can invert from the latent abundance scale to the
categorical abundances to predict abundance for unsampled cells and
also to predict abundance in the absence of land use transformation.

 With spatial data collection, we anticipate spatial pattern
in abundance and thus introduce spatial structure into our modeling.
That is, causal ecological explanations such as localized dispersal, as
well as omitted (unobserved) explanatory variables with spatial pattern
such as local smoothness of geological or topographic features, suggest
that, at sufficiently high resolution, abundance of a species at one
location will be associated with its abundance at neighboring locations
[\citet{Verhoef01}]. Moreover, through spatial modeling, we can provide
spatial adjustment to cells that have not been sampled, accommodating
the gaps in sampling and irregular sampling intensity mentioned above.
In particular, we create a latent \textit{process} model through a
trivariate spatial process specification, with truncated support, to
capture potential abundance, land transformation-adjusted abundance,
and measurement error-adjusted abundance. Since our environmental
information is available at grid cell level, we use Markov random field
(MRF) models [\citet{Besag74}, \citet{Banerjee04}] to capture spatial dependence
and to facilitate computation. However, we work with a large landscape
of approximately $37{,}000$ grid cells which leads to very long run times
in model fitting and so we introduce a novel parallelized computing
strategy to expedite fitting.

 There have been other recent developments in modeling of
species abundances, some using Bayesian hierarchical models. First,
there has been some work on developing models that deal almost
exclusively with animal census data, including count data and
mark-recapture data [\citet{Royle07}, \citet{Conroy08}, \citet{Gorresen09}]. \citet{Potts06} provide an overview of abundance modeling, in fact, five
regression models (Poisson, negative binomial, quasi-Poisson, the
hurdle model, and the zero-inflated Poisson) fitted for one particular
plant example. These models focus on correcting observer error and bias
as well as under-detection (the species is present but not observed),
whence the ``true'' abundance is virtually always higher than observed
[\citet{Royle07}, \citet{Cressie09}].
We note some very recent work on working with ordinal species abundance
in plant data by \citet{Ibanez09}. This approach takes ordinally scored
abundances and uses an ordered logit hierarchical Bayes model to infer
potential abundances for species that are still spreading across the
landscape.

 The advantages of working within the Bayesian framework with
Markov chain Monte Carlo (MCMC) model fitting are familiar by now---full inference about arbitrary unknowns, that is, functions of model
parameters and predictions, can be achieved through their posterior
distributions, and uncertainty can be quantified exactly rather than
through asymptotics. In this application we work with the disaggregated
data at the level of individual species and sites to present spatially
resolved abundance ``surfaces'' and to capture uncertainty in model
parameters. Doing this turns out to be more difficult than might be
expected, as we reveal in our model development section. The key
modeling issues center on careful articulation of the definition of
events and associated probabilities, the misalignment between the
sampling for abundance (at the relatively small sampling sites) and the
available environmental data layers (at a scale of minute by minute
grid cells, roughly 1.55~km${}\times{}$1.85~km over the region), the
sparseness of observations in terms of the entire landscape (with
uneven sampling intensity including many ``holes''), the occurrence of
considerable human intervention with regard to land use across the
landscape (``transformation''), and the need for spatially explicit
modeling.

 The format of the paper is as follows. Section~\ref{sec2}
describes the motivating data set. Section~\ref{sec3} develops the
multi-level abundance model. Section~\ref{sec4} details the
computational and inference issues. In Section~\ref{sec5} we present an
analysis of the data from the Cape Floristic Region (CFR) and conclude
with some discussion and future extensions in Section~\ref{sec6}.

\section{Data description}\label{sec2}

\begin{figure}[b]

\includegraphics{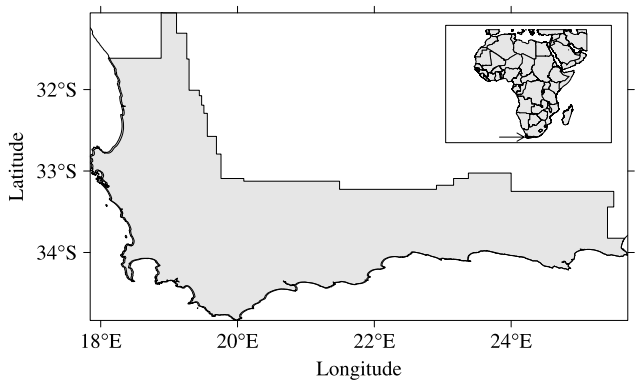}

\caption{Location of the Cape Floristic Region (CFR) of South Africa.
Inset shows the location of the CFR within the African Continent. The
$90{,}000$~km$^2$ region was divided into $36{,}907$ $1$-minute cells for modeling.}\label{fig1}
\end{figure}

 The focal area for this abundance study is the
Cape Floristic Kingdom or Region (CFR), the smallest of the world's six
floral kingdoms
(Figure~\ref{fig1}). As noted above, it encompasses a small region of
southwestern South Africa, about 90,000~km$^{2}$, including the Cape
of Good Hope, and is partitioned into 36,907 minute-by-minute grid
cells of equal area. It has long been recognized for high levels of
plant species diversity and
endemism across all spatial scales. The region includes about 9000
plant species, $69\%$
of which are found nowhere else [\citet{Goldblatt00}]. Globally, this is
one of the highest concentrations
of endemic plant species in the world. It is as diverse as many of
the world's tropical rain forests and apparently has the highest
density of
globally endangered plant species [\citet{Rebelo02}].
The plant diversity in the CFR is concentrated in relatively few
groups, such as the icon
flowering plant family of South Africa, the Proteaceae. We focus on
this family
because the data on
species distribution and abundance patterns are sufficiently rich and
detailed to allow complex modeling.
The Proteaceae have also shown a remarkable level of speciation, with
about 400 species
across Africa, of which 330 species are $99\%$ restricted to the CFR.
Of those 330 species,
at least 152 are listed as ``threatened'' with extinction by the
International Union for
the Conservation of Nature. Proteaceae have been unusually well sampled
across the region by the Protea Atlas Project of the South African
National Biodiversity Institute [\citet{Rebelo01}]. Data
were collected at \textit{record localities}: relatively uniform,
geo-referenced areas typically 50
to 100 m in diameter. In addition to the presence (or absence) at the
locality of protea
species, abundance of each species along with selected environmental
and species-level
information were also tallied [\citet{Rebelo91}]. To date, some 60,000
localities have been
recorded (including null sites), with a total of about 250,000 species
counts from among
some 375 proteas [\citet{Rebelo06a}].

 Abundance is given for a sampling locality. Evidently, there
is no notion of abundance at a point; however, with roughly $60{,}000$
sites sampled over the entire CFR, the relative scale of the Protea
Atlas observations is small enough when compared to our areal units to
be considered as ``points.'' In the literature, abundance is sometimes
measured as percent cover [\citet{Mueller03}]. In our data set,
abundance is recorded as an ordinal categorical classification of the
count for the species with four categories: category 0$:$ none
observed, category 1$:$ 1--10 observed, category 2$:$ 11--100
observed, category 3$:$ $>$100 observed. Such categorization is fast
and efficient for studying many species and many sampling locations but
is certainly at risk for measurement error in the form of
misclassification. Additionally, a large number of cells were not
sampled at all. In fact, only 10,158, that is, $28\%$, were sampled
at one or more sites. Even among cells sampled, some have just one or
two sites while others have more than 100, reflecting the irregular and
opportunistic nature of the sampling rather than an experimentally
designed sampling plan.

\begin{figure}[b]

\includegraphics{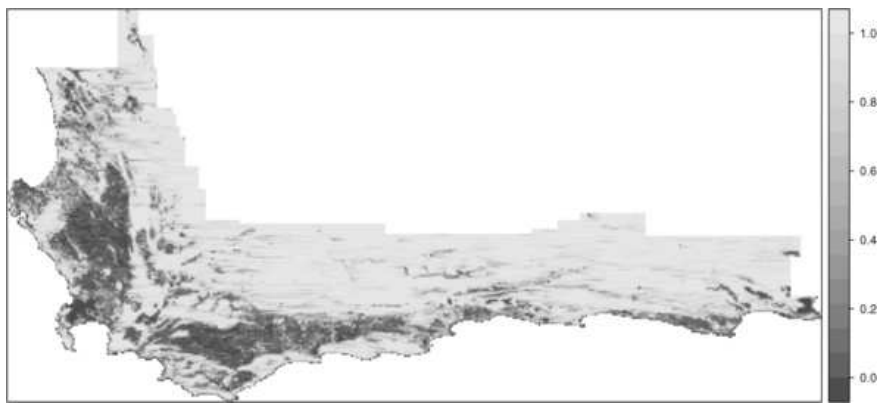}

\caption{Proportion of untransformed land inside the CFR. Most of the
transformation is due to agriculture, but includes dense stands of
alien invasive species.}
\label{fig2}
\end{figure}

 Turning to the covariates, in \citeauthor{Gelfand05a} (\citeyear{Gelfand05a,Gelfand05b})
$16$ explanatory environmental variables were studied, capturing
climate, soil, and topographic features (further detail is provided
there). Here, we confine ourselves to the six most significant
variables from that study, which are Evapotranspiration (APAN.MEAN),
July (winter) minimum temperature (MIN07), January (summer) maximum
temperature (MAX01), mean annual precipitation (MEAN.AN.PR), summer
soil moisture days (SUMSMD), and soil fertility (FERT1).
Transformed areas (by agriculture, reforestation, alien plant
infestation, and urbanization) were
obtained as a GIS data layer from R. Cowling (private communication).
Figure~\ref{fig2} shows the pattern of transformation across the CFR.
Approximately $1/3$ of the Cape has been transformed, mainly in the
lowlands on more fertile soils
where rainfall is adequate [\citet{Rouget03}]. Most of the
transformation outside of
these areas, on the infertile mountains, is due to dense alien invader
species, which are
currently a major threat to Fynbos vegetation and, in particular, to
the Proteaceae.

\section{Multi-level latent abundance modeling}\label{sec3}

In Section~\ref{subsec31} we briefly review the earlier work
on hierarchical modeling for presence/absence data, presented in
\citeauthor{Gelfand05a} (\citeyear{Gelfand05a,Gelfand05b}), in order to reveal how we have generalized it
for the abundance problem. Section~\ref{subsec32} develops our proposed
probability model for the categorical abundance data. In Section~\ref{subsec33} discrete probability distributions are replaced using latent
continuous variables. In Section~\ref{subsec34} we discuss bias issues
associated with modeling abundance data and, in particular, how they
affect our setting. Section~\ref{subsec35} deals with explicit model
details for the likelihood and posterior.

\subsection{Hierarchical presence/absence modeling}\label{subsec31}

 In \citeauthor{Gelfand05a} (\citeyear{Gelfand05a},\break \citeyear{Gelfand05b}) the authors model at the
scale of the grid cells in the CFR and provide a block averaged binary
process presence/absence model at this scale. In particular, let $A_i
\subset\mathbb{R}^2$ denote the geographical region corresponding to
the $i$th grid cell and $X_{i}^{(k)}$ the event that a randomly
selected location within $A_i$ is\vspace*{-1.5pt}
suitable (1) or unsuitable (0) for species $k$. Set $P(X_{i}^{(k)} =1)
= p_{i}^{(k)}$. Then, $p_{i}^{(k)}$ is
conceptualized by letting $\lambda^{(k)}(s)$ be a binary process over
the region
indicating the suitability (1) or not (0) of location $s$ for species
$k$ and taking $p_{i}^{(k)}$ to be the
block average of this process over unit $i$. That is,
\begin{equation}\label{eq311}
p_{i}^{(k)} = \frac{1}{|A_{i}|} \int_{A_i} \lambda^{(k)}(s)\, ds
= \frac{1}{|A_{i}|} \int_{A_i} \mathbf{1}\bigl(\lambda^{(k)}(s) =1\bigr)\,ds,
\end{equation}
where $|A_{i}|$ denotes the area of $A_i$. From Equation (\ref{eq311}),
the interpretation is that the more
locations within $A_i$ with $\lambda^{(k)}(s) = 1$, the more suitable
$A_i$ is for species $k$, that is, the
greater the chance of potential presence in $A_i$. The collection of
$p_{i}^{(k)}$'s over the $A_i$ is viewed as the potential distribution
of species $k$.\vspace*{-2pt}

 Let $V_{i}^{(k)}$ denote the event that a randomly selected
location in $A_i$ is suitable for
species $k$ in the presence of transformation of the landscape. Let
$T(s)$ be an indicator
process indicating whether location $s$ is transformed ($1$) or not (0).
Then, at $s$, both $T(s) = 0$ (availability) and $\lambda^{(k)}(s) = 1$
(suitability) are needed in order that location $s$ is suitable
under transformation. Therefore,
\begin{equation}\label{eq312}
P\bigl(V_{i}^{ (k)}=1\bigr) = \frac{1}{|A_{i}|} \int_{A_i} \mathbf{1}\bigl(T(s)=0\bigr)
\mathbf{1}\bigl(\lambda^{(k)}(s)=1\bigr)\, ds.
\end{equation}
If, for each pixel, availability is uncorrelated with suitability, then
Equation (\ref{eq312}) simplifies to $P(V_{i}^{(k)}= 1) = u_{i}p_{i}^{(k)}$,
where $u_{i}$ denotes the proportion of area in $A_i$ which is
untransformed, $0 \leq u_{i} \leq1$.

 Next, assume that $A_i$ has been visited $n_{i}$ times in
untransformed areas within the cell. Further, let $y_{ij}^{ (k)}$
be the observed presence/absence status of the $k$th species at the
$j$th sampling location within the $i$th unit. The $ y_{ij}^{ (k)} |V_{i}^{(k)}=1$ are\vspace*{-3pt}
modeled as i.i.d. Bernoulli trials with success\vspace*{-2pt}
probability $q_{i}^{(k)}$, that is, for a randomly selected location in\vspace*{-2pt}
$A_i$, $q_{i}^{(k)}$ is the probability of species $k$ being present given the location is
both suitable and available. Of course, given $V_{i}{ (k)} = 0$,\vspace*{-3pt}
$y_{ij}^{ (k)} = 0$ with probability 1. Then, we have that
$P(y_{ij}^{(k)}=1) = q_{i}^{(k)} u_{i} p_{i}^{(k)}$.\vspace*{-3pt}
\citeauthor{Gelfand05a} (\citeyear{Gelfand05a,Gelfand05b}) model the $p_{i}^{(k)}$ and $q_{i}^{(k)}$ using
logistic regressions. In fact, they use environmental variables and
spatial random effects to model the $p_{i}^{(k)}$'s, the probabilities
of potential presence, and, to facilitate identifiability of
parameters, use species level attributes to model the $q_{i}^{(k)}$'s.

\subsection{Probability model for categorical abundance}\label{subsec32}

 We first define what \textit{categorical abundance} means at an
areal scale using the four ordinal categories from Section~\ref{sec2}.
Suppressing the species index, let $X_i$ denote the classification for
a randomly selected location in $A_i$ and define $p_{ih}$ $=$ $P(X_i =
h)$ for $h=0,1,2,3$. If $\lambda(s)$ is a four-colored process taking
values $0,1,2,3$, then $p_{ih} = \frac{1}{|A_i|} \int_{A_i} 1(\lambda
(s)=h)\, ds$. That is, $p_{ih}$ is the proportion of area within $A_i$
with color $h$, equivalently, the proportion in abundance class $h$.
The $p_{ih}$ denote the potential abundance probabilities, that is, in
the absence of transformation.

 We describe land transformation percentage $(1-u_i)$ as a
block average of a binary availability process $T(s)$ over $A_i$. It
can also be interpreted as the probability that a randomly selected
site within $A_i$ is transformed. At a transformed location abundance
\textit{must} be $0$. Thus, as in Equation (\ref{eq312}), in the presence
of transformation, we revise $p_{ih}$ to $P_T(X_i=h)= \frac{1}{|A_i|}
\int_{A_i} 1(T(s)=0)1(\lambda(s)=h)\, ds$. Under independence of
abundance and land transformation, we obtain $P_T(X_i=h)=u_ip_{ih}$.
The $u_i p_{ih}$ denote the transformed abundance probabilities for $h
\neq0$. The probability of abundance class $0$ under transformation is
evidently $1-u_i + u_i p_{i0}$. Let $r_{ih}$ denote the abundance class
probabilities in the presence of transformation.

 Finally, suppose there is an observed categorical abundance
at location $j$ within $A_i$, say, $y_{ij}$. There is an associated
conceptual $\lambda_{ij}$ and an observed $T_{ij}$. Then, $\lambda_{ij}
\neq\lambda_{ij} T_{ij}$ if there has been transformation degradation
at location $j$, unless $\lambda_{ij}=0$. Furthermore, if there has
been a misclassification error at $j$, $y_{ij} \neq\lambda_{ij}
T_{ij}$ unless $\lambda_{ij}=0$. Let $q_{ih}$ denote the abundance
class probabilities associated with the observed abundances.
In Section~\ref{subsec33} we specify a latent trivariate continuous abundance
model that produces the $p$'s, $r$'s, and $q$'s by integrating over
appropriate intervals. This latent model can be viewed as the \textit{process model} for our setting.

\begin{figure}

\includegraphics{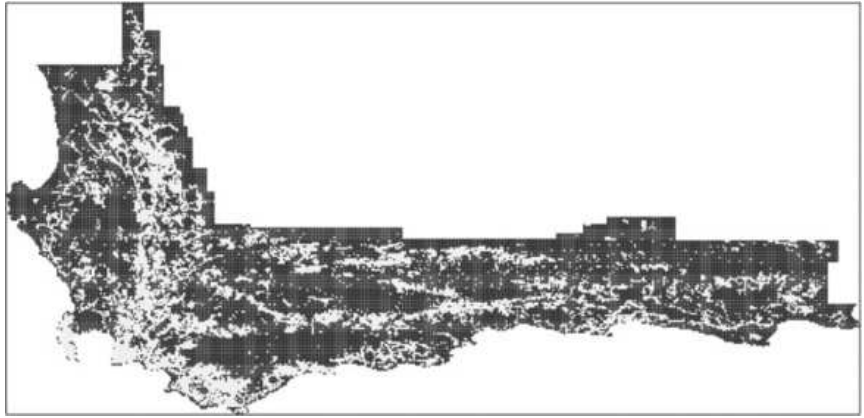}

\caption{Cells within the CFR that have at least one observation from
the Protea Atlas data set are shown in light grey, while cells with no
observations are shown in dark grey.}
\label{fig3}
\end{figure}

 The data set consists of observed abundances across several
sampling sites within the CFR. Let $D$ denote our CFR study domain so
$D$ is divided into $I=36{,}907$ grid cells of equal area. For each cell
$i=1,2,3,\ldots,I$, we are given information on $p$ covariates as $v_i =
(v_{i1},v_{i2},\ldots,v_{ip})$. Within $A_i$, the abundance category of a
species was recorded at each of $n_i$ sampling sites. For many cells
$n_i > 1$. For site $j$ in $A_i$ we observe $y_{ij}$ as a multinomial
trial, that is, $y_{ij} \stackrel{\mathrm{i.i.d.}}{\sim}\operatorname{mult}(\{q_{ih}\}),
j=1,2,\ldots,n_i$. We have a large number of unsampled cells, that is,
$n_i=0$. In fact, out of 36,907 cells, only $m=10{,}158$ ($28\%$) were
sampled at one or more sites. Figure~\ref{fig3} indicates locations of
sampled cells. For the unsampled cells there are no $y_{ij}$'s in the
data set. Hence, the inference problem involves estimation of
probabilities over the observed cells as well as \textit{prediction}
over the unsampled region. Prediction of a categorical response
distribution for unsampled locations in a point level model was
discussed in \citet{Oliveira00} and \citet{Higgs09}. In our areal setup
with only areal level $v$'s, we address this problem with a MRF model,
as described in Section~\ref{subsec33}. Again, we seek to infer about
the $p$'s, $r$'s, and $q$'s given the observed $y$'s for a subset of
cells and $v$'s known for all cells.
%

\subsection{Latent continuous abundances}\label{subsec33}

 It is now common to model the probability mass function of a
scalar \textit{ordinal} categorical variable through an underlying
univariate continuous distribution. In a more general setup, \citet{Loch97} and \citet{Armstrong03} used latent random vectors to define
the categorical probabilities in terms of these vectors taking values
within a specific set. In a similar spirit, corresponding to an
observed abundance category variable $y_{ij}$, we introduce a
continuous latent variable $z_{O,ij}$ such that
\[
y_{ij} = \sum_{h=0}^{3}h 1(\alpha_{h-1} < z_{ij} < \alpha_h),
\]
 where $\alpha= (\alpha_{-1} = -\infty,\alpha_0 = 0,\alpha
_1,\alpha_2,\alpha_3=\infty)$ are an increasing sequence of cut points.
For identifiability and without loss of generality, we can set $\alpha
_0=0$ and interpret $z_{O,ij}<0$ as an absence, $z_{O,ij}>0$ as a
presence. We have $P(y_{ij}=h)= q_{ih}=P(z_{O,ij} \in(\alpha
_{h-1},\alpha_{h}))$. So $q_{ih}$ will be determined by the probability
model specified for the $z_{O,ij}$'s. We will introduce spatial
dependence between $z_{O,ij}$'s below but, for now, to simplify
notation, we drop the subscript.

 A simple model would put a Gaussian distribution on these
latent $z_{O}$'s whose means are linear functions of the associated
$v$'s. This would provide a routine categorical regression model but
ignores known land transformation and potential measurement/ecological
error in the recorded abundance categories. Instead, we introduce
$z_{P,ij}$ to provide the $p_{ij}$'s and $z_{T,ij}$ to provide the
$r_{ij}$'s. We need a joint distribution to relate the $z_{P}$,
$z_{T}$, and $z_{O}$. From a process perspective in terms of the
proposed degradation, it seems natural to specify this distribution in
the form $f(z_{P}) f(z_{T}|z_{P}) f(z_{O}|z_{T})$. Since
($z_P,z_T,z_O$) capture the sequential degradation of an associated
categorical abundance distribution, we need to use the same set of
$\alpha$'s to produce meaningful ($p,r,q$) respectively. Now, we
propose (and clarify) the following dependence structure. Define $c(\mu
) = \mu- \frac{\phi(\mu)}{\bar\Phi(\mu)}$, where $\phi(\cdot)$ and
$\Phi(\cdot)$ are the standard normal p.d.f. and c.d.f. respectively. Note
that $c(\mu)= E(V|V \sim N(\mu,1), V < 0)$ so $c(\mu)\leq \min(0,\mu)$
for all $\mu\in\mathbb{R}$. Let
%
\begin{eqnarray}\label{eq331}
P(y=h|z_{O},u) &=& 1({\alpha_{h-1}\leq z_{O}\leq\alpha_{h}}); \qquad  0\leq h \leq3, \nonumber\\
f(z_{O}|z_{T}) &=& \phi(z_{O};z_{T},1)1_{z_{T} \geq0} + \delta_{z_{T}}1_{z_{T} \leq0}, \nonumber\\[-8pt]\\[-8pt]
f(z_{T}|z_{P},u) &\sim& u\delta_{z_{P}} + (1-u)\delta_{c(z_{P})},\nonumber\\
f(z_{P}|v,\beta,\tau^2) &=& \phi(z_{P};v^T \beta + \theta,1).\nonumber
\end{eqnarray}

 Again, the conditional modeling above is motivated by the
degradation perspective. To model the latent $z_{P}$ surface, we use
the covariate information, that is, climate and soil features that are
believed to influence the abundance distribution of different species
in different ways. We also add a spatial random effect ($\theta$) in
the mean function to account for spatial association that may arise
from factors, apart from included covariates, that may have a spatial
pattern. The covariate effects $\beta$ as well as the spatial random
effects $\theta$ are species-specific. Variances are fixed at 1 for
identifiability (see Section~\ref{subsec35}). Since we are working at
areal scale, we assign each cell a single $\theta$ with the prior on
$\theta_{1,2,\ldots,I}$ specified using a Gaussian Markov random field
(MRF) [\citet{Besag74}] with first-order adjacency proximities. See
\citet{Banerjee04} for details as well as further references.

 Next, the $z_P$ surface is degraded by land transformation. A
random location inside $A_i$ is untransformed with probability $u_i$.
Then, $z_T=z_P$, that is, a degenerate distribution at $z_P$ given
$z_P$. If it is transformed, the degradation occurs so that the $z_T$
corresponds to the zero abundance category. For simplicity (with
further discussion below), we make this a degenerate distribution at
$c(z_P)<0$, whence $z_T|z_P,u$ becomes a two point distribution as
above. Again, transformation is equivalent to absence and since $\alpha
_0=0$ is the upper threshold for that classification, we need $z_T <0$
for a transformed location. When a cell is completely transformed, from
Equation (\ref{eq331}) we have $z_T <0$ w.p. $1$. For $u=1$ (complete
availability) $z_T$ and $z_P$ are the same. For any $0<u<1$, we get
$E(z_T|z_P) = uz_P + (1-u)c(z_P).$

 Also, since $c(x)<x$, $E(z_T|z_P) \leq z_P$, which is
essential in the sense that transformation can only degrade abundance
[and clarifies our choice for $c(\cdot)$]. Posterior summaries of $z_T$
measure the prevailing abundance under transformation within the CFR.
[In Appendix~\ref{A1} we show that $|E(z_T)|<\infty$.] The two-point
mixture distribution also implies the probability of abundance class
$0$, $ P(z_T<0)\geq(1-u)$, that is, no matter how large the potential
abundance is within a cell, for any $u<1$ there is a positive
probability that transformed abundance may fall below $0$ at a random
location within the cell. Other choices for the $z_T|z_P$ specification
besides a point mass at $c(z_{P})$ include putting a point mass at some
arbitrary point $c<0$, or using a truncated normal $z_T|z_P$ on $\mathbb{R}^{-}$. In the first case, it is not ensured whether $z_T \leq z_P$
(it depends on whether there are cells with $z_P < c$), while the
second choice adds complication for no benefit, is less interpretable,
and does not ensure $z_T<z_P$ with probability $1$. Also, in Section~\ref{sec4} we show that, in terms of fitting the model, the
specification used in Equation (\ref{eq331}) is the same as using a
truncated normal distribution for land transformation.

Next, we modify the \{$z_{T}$\} surface to produce \{$z_{O}$\}.
With regard to measurement error, the recorded category of abundance
at a particular location can be different from the prevailing category
due to inaccuracy in field assessment of species quantity.
However, we assume that when the potential abundance was zero, one
could not record a nonzero abundance category for it [no false
positives, see \citet{Royle06} in this regard]. This puts a directional
constraint on the effect of noise. A specification for $f(z_{T}|z_{O})$
which is coherent with this restriction has, with $z_T>0$ (i.e., a
presence), $z_O|z_T \sim N(z_T,1)$. This is a usual measurement error
model (\textit{MEM}) specification. For a site with no presence $z_T< 0$, our
assumption says there cannot be any measurement error, thus, in
Equation (\ref{eq331}), for simplicity, we set $z_O$ to be the same as
$z_T$. Again, other choices of $z_O|z_T$ can be considered for the $z_T
< 0$ event, but they will not have any impact on estimation of the
$z_P$ surface, as we clarify in Section~\ref{sec4}.
This sequential dependence structure, $z_P\rightarrow z_T \rightarrow
z_O$, implies that if $z_P<0$ so is $z_T$ and $z_O$. Hence, if a site
is not suitable for a species, at no intermediate stage of the model
can the site have any positive probability of species occurrence. A
change in category between \textit{actual} and \textit{observed} arises
when the noise pushes $z_{T}$ to the other side of some cut point to
produce $z_{O}$. And, because of the truncation structure, that shift
cannot happen from the left of $\alpha_0=0$ to the right.

 An alternative way to jointly model $(z_{O},z_{T})$ could use
a bivariate normal distribution with support truncated to $\mathbb{R}^2
- [0, \infty) \times(-\infty,0]$. However, this specification fails to
produce an $f(z_{O}|z_{T})$ which match our intuition about how the
degradation took place. Also, from the distributional perspective, the
truncated normal redistributes the mass contained inside the left-out
region uniformly across the support, whereas the specification in
Equation (\ref{eq331}) shifts the mass only to $(z_{O}<0)$, which is
more in agreement with modeling a data set such as ours where we have
an inflated number of reported zero abundances.

 The simple dependence structure for $z_T|z_P$ allows us to
marginalize over $z_T$ and work with $z_P$ and $z_O|z_P$ as our joint
latent distribution. We have
%
\begin{eqnarray}\label{eq332}
f(z_{O}|z_{P}) =
\cases{
u  \phi(z_{O};z_{P},1) + (1-u)\delta_{c(z_{P})},  &\quad$z_P>0$, \cr
u  \delta_{z_P} + (1-u)\delta_{c(z_{P})},  &\quad$z_P<0$.
}
\end{eqnarray}
 Rewriting Equation (\ref{eq332}) in a simpler form, we get
%
\begin{eqnarray}\label{eq333}
f(z_{O}|z_{P}) &\sim& u  [ \phi(z_{O};z_{P},1)1_{z_{P} \geq0} +
\delta_{z_{P}}1_{z_{P} \leq0}  ] + (1-u)\delta_{c(z_{P})}.
\end{eqnarray}
 Moreover, Equation (\ref{eq333}) has a nice interpretation in
the sense that, first, it indicates whether the land is transformed or
not with probability $1-u$. If the land is transformed, it sets
observed abundance to be $z_O=c(z_P)<z_P$. In the case of available
land, if there is a potential presence, it allows for a MEM around
$z_P$; in the case of absence, it stays fixed at $z_P$. Since $z_O$ is
related to the observed data and $z_P$ is our surface of interest, the
marginalization removes one stage of hierarchy from our model fitting
and thus reduces correlation, yielding better behaved MCMC in model
fitting. Furthermore, we can retrieve the $z_{T}$ surface after the
fact since $f(z_T|z_O,z_P) \propto f(z_T|z_P)f(z_O|z_T)$.

\subsection{Measurement error and bias issues}\label{subsec34}

 In the \hyperref[sec1]{Introduction} we noted that measurement error and bias
typically occur with ecological survey data. It can manifest itself in
the form of detection error, spatial coverage bias [\citet{Royle07}],
and under-reporting of absences. How do these biases arise in our
modeling? Noteworthy points here are (i) the difference between
obtaining abundance as actual counts as opposed to through ordinal
classifications and (ii) what ``no abundance'' means across our
collection of grid cells.

 Nondetection bias (i.e., undetected individuals in a sampled
location) tends to be discussed more with regard to animal abundance
[\citet{Verhoef03}, \citet{Royle07}, \citet{Gorresen09}]. Using counts, evidently
observed abundance is at most true abundance; error can occur in only
one direction. With ordinal counts, the bias is still expected to
reflect under-reporting but, depending upon the categorical
definitions, will be much less frequent and need not be absolutely so.
For example, in our data set, plant population size is visually
estimated and an observation, especially of large populations, could
potentially have error in either direction. In our modeling, ``true''
abundance is not ``potential'' abundance. For us, one could envision
true abundance on the latent scale as a ``true'' transformed abundance,
say, $\tilde{z}_{T}$ with measurement error yielding $z_{O}$. Then, one
might insist that our measurement error model requires $z_{O} \leq
\tilde{z}_{T}$. Under our measurement error formulation using $z_{T}$,
we even allow $z_{O} > z_{P}$ to account for potential overestimation
of abundance. Evidently, since $y_{O}$ may occasionally be less than
the potential classification $y_{P}$ at that location, we may be
slightly underestimating potential abundance. We don't expect this to
be consequential and, in any event, with no knowledge about the
incidence of \textit{under-classification} in our setting, we have no
sensible way to correct for this bias.

 Turning to spatial coverage bias (i.e., individuals not
exposed to sampling will be missed), for us, with only $28\%$ of grid
cells sampled, we certainly are subject to this. However, the spatial
modeling we introduce helps in this regard. The mean of $z_{P,ij}$ is
$v_i^{T} \beta+ \theta_{i}$ regardless of whether we collected any
data in $A_i$. So, the regression is expected to find the appropriate
level for the cell and the spatial smoothing associated with the $\theta
_{i}$ is expected to provide suitable local adjustment. We could argue
that, if sampling of grid cells is random, this bias can be ignored.

 Perhaps the most difficult bias to address is the
under-sampling of absences. This bias counters the previous ones;
under-sampling of absences will tend to produce over-estimates of
potential abundance. In our setting, under-sampling of absences is
reflected in the decision-making that leads to only $28\%$ of cells
being sampled, that is, it is not a random $28\%$ that have been
sampled. Different from spatial coverage bias, in this context, the
ecologist expresses confidence that the species is not present in some
of the unsampled cells. If this is so and we were to set some
additional abundances to $0$, this would assert that these ``0''s are
not nondetects and would diminish potential abundance, opposite to the
case of nondetects. Of course, in the absence of actual data
collection, we would not see any of these $0$'s and would adopt
model-based inference regarding potential abundance for these cells. In
any event, with no explicit knowledge of how sampling sites were
chosen, we are unable to attempt correction for this bias. Possibly,
approaches to address the effects of preferential sampling [\citet{Diggle10}] could be attempted here.

\subsection{Likelihood and posterior distribution}\label{subsec35}

 The posterior distributions of interest, $p$ and $r$, will be
constructed in the post MCMC analysis (discussed in detail in
Section~\ref{subsec43}). From the conditional structure we first write
$P(y=c|z_{O},\alpha) = 1_{z_{O}\in(\alpha_{c-1},\alpha_{c})}$. So
the likelihood function for a single sample $y$ turns out to be
$L(y|z_{O},\alpha) = \prod_{k=0}^{3} 1({z_{O}\in(\alpha_{k-1},\alpha
_{k})})^{1(y=k)}$. Now $f(z_{O}|z_{P})$ can be written as in Equation
(\ref{eq333}).

 Again, we have $I$ cells with $n_i$ sampling sites within
$A_i$. For each $y_{ij}$ we introduce a corresponding $z_{O,ij,}$ and
hence a pair of $z_{T,ij},z_{P,ij}$, to represent the event happening
at the $j$th sampling site within $A_i$. We work directly with the
$z_O|z_P$ structure. Since we are interested in the areal level
abundance distribution and have covariates at areal resolution, we
assume for fixed $i$, $z_{P,ij}$ $\stackrel{\mathrm{i.i.d.}}{\sim}$ $N(\cdot ;
v_i^T \beta  + \theta_i,1)$. It is also assumed that the $z_{O,ij}$'s
are conditionally independent given the $z_{P,ij}$'s.

 Without loss of generality, re-index cells so that the first
$m$ of them are sampled and the last $I-m$ are not. The latter cells
have no contribution to the $y$ column and, hence, no associated
$z_{O}$ appears in the likelihood. Using a nonspatial model, we would
work with a posterior on the domain of sampled cells only. But assuming
a CAR prior structure with adjacency proximity matrix $W$ for the
$\theta$ over the whole domain enables us to learn about $z_P$ for
unsampled cells. In summary, the posterior distribution takes the
following form, up to proportionality, with $\Theta= (\bolds{\alpha
},\bolds{\beta},\bolds{\theta})$:
%
\begin{eqnarray}\label{eq351}
\hspace*{35pt}\pi(\mathbf{z}_{P},\mathbf{z}_{O},\Theta|\mathbf{y,v,u})
&\propto&\prod_{i=1}^m \prod_{j=1}^{n_i} L(y_{ij}|z_{O,ij},\bolds{\alpha})f(z_{O,ij}|z_{P,ij})f(z_{P,ij}|v_i,\Theta)\nonumber
\\
&&{}\times\pi(\Theta), \nonumber\\[-8pt]\\[-8pt]
\pi(\Theta)
&=& \pi(\bolds{\alpha})\pi(\bolds{\beta})\pi(\bolds{\theta}),\nonumber \\
\pi(\theta_{1,2,\ldots,I})
&=& \operatorname{CAR}(\eta_0,W).\nonumber
\end{eqnarray}

 We turn to the identifiability of the set of parameters under
the hierarchical dependent latent structure. First, with a latent
continuous process yielding an ordinal categorical variable, the mean
and scale of the distribution can be identified only up to their ratio.
In Equation (\ref{eq331}) the dependence across $(z_p,z_T,z_O)$ is
specified through conditional means. Hence, all Gaussian distributions
there are specified with standard deviation $1$. Four categories of
abundance allow three free probabilities and the corresponding four
latent surfaces will also have $3$ degrees of freedom. As noted above,
we set $\alpha_{0}=0$ with $\alpha_1,\alpha_2$ as free parameters.

 We also need to ensure that all three $z$ surfaces can be
distinguished from each other. Since transformation percentage $1-u$ is
given a priori, it is straightforward to separate $z_P$ and $z_T$. We
turn to the joint distribution for $z_P, z_O$ given as $z_O|z_P \sim
N(z_P,1), z_P \sim N(v\beta+\theta,1)$. With fixed variances and no
constraint on measurement error, there would be no need to bring in
$z_P$; it is redundant, there is no way to distinguish between $z_P$
and $z_O$, and one can use the marginal $z_O\sim N(v\beta+\theta,2)$.
Now the constraint comes into the picture; it makes the $z_O$ surface
non-Gaussian though the $z_P$ surface is.
The greater the measurement error, the more departure from Gaussianity
in the marginal distribution of $z_O$. Again, the measurement error cannot be estimated on any absolute scale, since the latent $z$ scales are
fixed for identifiability. It will be controlled by parameters like
$\beta$ and~$\theta$. To compare the relative effect of measurement
error across different species, under fixed scale parameters,
$P(z_P<0)$ is a candidate but other model features can be informative
as well.

\begin{figure}

\includegraphics{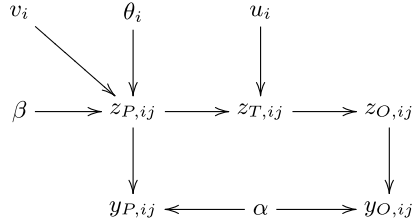}

\caption{Graphical model for latent abundance specification at site
$j$ within cell $A_i$. $z$'s denote latent abundance processes,
observed (O), transformed (T), and potential (P); $y$'s denote
interval-censored abundances, observed (O), and potential (P); $u$ is
proportion of land untransformed, $v$'s are covariates, $\beta$'s are
regression coefficients, $\alpha$'s are cut points for $z$ scale, and
$\theta$'s are spatial random effects.}
\label{fig4}
\end{figure}

 Finally, the full model specification, described in Equation
(\ref{eq331}), can be represented through a graphical model, shown in
Figure~\ref{fig4}.
%

\section{Posterior computation and inference}\label{sec4}

 Here, we describe how to design a computationally efficient
MCMC algorithm for the model. We then discuss how to summarize the
posterior samples to estimate important model features.

\subsection{Sampling}\label{subsec41}

 Introduction of latent layers, although increasing the
parameter dimension in the model, makes the posterior full conditionals
standard and easy to sample from. Our goal is to efficiently estimate
components of $\Theta$ which control potential abundance. We rewrite
Equation (\ref{eq333}) as follows:
%
\begin{eqnarray}\label{eq411}
f(z_{O}|z_{P}) &=& u\phi(z_{O};z_{P},1)1_{z_{P} \geq0}
+  \bigl[ u\delta_{z_{P}}+(1-u)\delta_{c(z_P)} \bigr]1_{z_{P} \leq0} \nonumber\\[-8pt]\\[-8pt]
&&{}  +   (1-u)\delta_{c(z_{P})}1_{z_P>0},\nonumber
\end{eqnarray}
and work with Equation (\ref{eq411}) to implement the computation for
the model fitting.

 We start with updating all $z_{O},z_{P}$ using $(\Theta
^{(t)};\mathbf{y,v,u})$ and then drawing components of $\Theta$ from
their respective posterior full conditionals based on $z_{P,(t+1)},$
$z_{O,(t+1)}$. Given the draw from $z_{P}$, sampling the components of
$\Theta$ is standard as in almost any spatial regression analysis (see
Appendix~\ref{A3}). For the set of $\theta$'s, after sampling them
sequentially, we need to ``center them on the fly'' [\citet{Besag95}, \citet{Gelfand99}]. The more challenging part is to update $z_O,z_P|\Theta$.
In \citet{Albert93} the latent variables were sampled in the MCMC from
mutually independent truncated Gaussian full conditionals, with the
support determined by the corresponding classification. For our model,
the posterior full conditional for any $z_{O}$ is
\begin{eqnarray*}
\pi(z_{O}|z_{P},y,u) \propto   f(z_{O}|z_{P}) 1\bigl({z_{O}\in(\alpha
_{y-1},\alpha_{y})}\bigr).
\end{eqnarray*}

 We take two different strategies to update $z_{P},z_{O}$
depending on the observed~$y$. For any site with nonzero $y$ we have
(with $\alpha_0=0$) $f(z_{O},z_{P}|y > 0,u)\propto\phi
(z_{O}|z_{P})1_{z_{O}\in(\alpha_{y-1},\alpha_{y})}\phi(z_{P})$, which
amounts\vspace*{-1.5pt} to sampling first a univariate normal
$z_{O,(t+1)}|(z_{P,(t)},y,\alpha)$ truncated within $(\alpha
_{y-1}^{(t)},\alpha_{y}^{(t)})$ and then from $z_{P,(t+1)}|$
$z_{O,(t+1)},\Theta^{(t)}$ which is also Gaussian [with location
$(z_{O,(t+1)}+{\mu}^{(t)})/2$ and scale $\sqrt{1/2}$ where $\mu
^{(t)}=v^T\beta^{(t)}+\theta^{(t)}]$. For a site with observed $y=0$
the case is more complicated, with details provided in Appendix~\ref{A2}. All of the sampling distributions required in MCMC are listed in
Appendix~\ref{A3}.

\subsection{Computational efficiency}\label{subsec42}

 The algorithm described above is computationally demanding as
we have two latent variables to sample at each sampling site and one
spatial parameter for each of the grid cells. However, since $z_P,z_O$
are independent across cells given $\Theta$, we can update them all at
once. The problematic part is sampling the spatial effects, with
approximately 37,000 grid cells. To handle this issue, we used a
parallelization method where $D$ is divided into disjoint and
exhaustive subregions $D_1,D_2,\ldots,D_L$ along with a resultant set of
\textit{boundary} cells $B$ arising through the CAR proximity matrix.
Thus, once $\theta_{B}$ is updated conditional on the rest, then $\theta
_{D_1},\theta_{D_2},\ldots,\theta_{D_L}$ given $\theta_{B}$ can be updated
in parallel.

\begin{figure}

\includegraphics{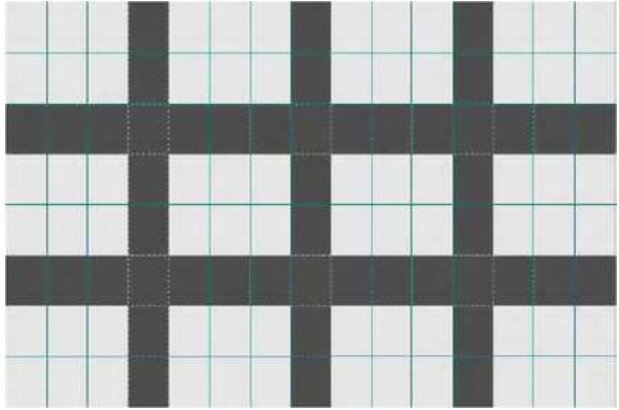}

\caption{An example grid to illustrate parallel CAR implementation.
Normal sequential updating would have required $120$ steps in each
iteration. By dividing the rectangle into $12$ segments of $6$ cells each
with $48$ boundary cells (shown in dark grey), each segment can be
updated independently (conditional on the boundaries). In this example
the parallelization results in only $54$ updating steps.}
\label{fig5}
\end{figure}

 This algorithm is illustrated in Figure~\ref{fig5}, where we
have a $15 \times8$ rectangular region with an adjacency structure $W$
which puts weight only on the cells sharing a common boundary.
Sequential updating would have required $120$ steps. We constructed a
set of $48$ boundary cells $B$ (the dark cells in Figure~\ref{fig5}).
It divides the rectangle into $12$ segments of $6$ cells each, so that
conditional on $\theta_B$, those segments can be updated independently
of each other so we need only $54$ updating steps. This is only an
illustrative example, but for large regions, this may significantly
improve the run time. However, the time required for communication and
data assimilation is an issue for this parallelization method. With
increasing $L$, although the time required for the sequential updating
within each $D_i$ goes down, the size of $B$ increases as does the
amount of communication required within the parallel architecture. So a
trade-off must be determined for choosing $L$; in our setting $L=11$
worked well.

\subsection{Posterior summaries}\label{subsec43}

 There are several ways to summarize inference about the $p$
and $r$ distributions. According to our model, for $A_i$, $p_{ih} = \Phi(\alpha_h -v_i^T\beta-\theta_i)-\Phi(\alpha_{h-1} -v_i^T\beta
-\theta_i)$. Posterior samples of $\beta,\theta,\tau^2$ enable us to
compute samples of the $p_i$. A posterior sample of $r_i$ can be
constructed using the relation $r_{i} \equiv
(1-u_i+u_ip_{io},u_ip_{i1},u_ip_{i2},u_ip_{i3})$. Additionally, we can
calculate the mean as well as the uncertainty from these samples,
enabling maps for transformed abundance ($r$) and potential ($p$)
abundance. For each of $p_i$ and $r_i$, we have $4$ submaps, one for
each abundance category. This is useful in terms of assessing high and
low abundance regions for the species. The $\beta$'s provide (up to
fixed scale) the effect of a particular climate or soil-related factor
on the abundance of a particular species. Comparison of the $p$ and $r$
maps informs about the effect of land transformation. One may also be
interested in capturing $p$ or $r$ through a single summary feature
rather than all $4$ categorical probabilities. Grouped mean abundance
(expectation with respect to the $p$ or $r$ distribution) can be used
with suitable categorical midpoints. We note that the posterior
inference can also be summarized on the latent scale using posterior
samples of the $z$'s. However, working on the $z$ scale can only
provide relative comparison.

\section{Data analysis}\label{sec5}

\begin{table}[b]
\caption{Posterior summaries for covariate effects (mean and $95$\% c.i. width)}\label{table1}
\begin{tabular*}{\textwidth}{@{\extracolsep{4in minus 4in}}ld{2.5}d{2.5}d{2.5}d{2.5}d{2.5}d{2.5}@{}}
\hline
\multicolumn{1}{@{}l}{\textbf{Species}}
&\multicolumn{1}{c}{\textbf{Apan.mean}}
&\multicolumn{1}{c}{\textbf{Max}$\bolds{01}$}
&\multicolumn{1}{c}{\textbf{Min}$\bolds{07}$}
&\multicolumn{1}{c}{\textbf{Mean.an.pr}}
&\multicolumn{1}{c}{\textbf{Sumsmd}}
&\multicolumn{1}{c@{}}{\textbf{Fert}$\bolds{1}$}\\
\hline
PRPUNC &1.2275 &-0.9436 &-0.8248 &0.2439 &0.1834 &0.0306\\
&(0.3809) &(0.2768) &(0.1143) &(0.1158) &(0.2006) &(0.1089)\\[3pt]
PRREPE &0.6825 &-0.4512 &-0.0864 &0.1753 &-0.2958 &0.0566\\
&(0.1710) &(0.1179) &(0.0612) &(0.0673) &(0.0996) &(0.0455)\\
\hline
\end{tabular*}
\end{table}

 We have implemented the described model on abundance data for
several different plant species over the whole \textit{CFR}. We centered and
scaled all the $v$'s before using them in the model. As priors we used
$\pi(\bolds{\alpha})\equiv1$, $\pi(\beta)=N(0,\phi I)$ with large $\phi
=100$. For $\theta$, we used $\eta_0^2 =0.1$ and $W$ to be a binary
matrix with $w(i,j)=1$ \textit{iff} $d(i,j)<0.30$. The threshold $0.30$
was used to provide an $8$ nearest neighbor structure for most of the
cells. However, for boundary cells, the number of neighbors varies from
$3$ to $6$. The parallelization algorithm was implemented inside R
(\url{http://www.r-project.org}) using $l=11$. The
run time for an individual species was about $9000$ iterations$/$day. The
outputs presented below are created by first running $12500$ iterations
of MCMC, discarding the initial $7500$ samples, and thinning the rest
at every fifth sample. The $\beta$'s were quick to converge, but the
$\alpha$ sequences were highly autocorrelated and moved more slowly in
the space.

 Here we consider two species, \textit{Protea punctata}
(PRPUNC) and \textit{Protea repens} (PRREPE). A summary of the model
output is presented through the following table and diagrams. Table~\ref{table1} provides the mean covariate effects for both species along
with the $95$\% equal tail credible interval width (in parentheses).
Considering $95$\% equal tail credible interval, all the covariate
effects are significant \textit{except} Fert1 for \textit{P.
punctata}.

\begin{figure}

\includegraphics{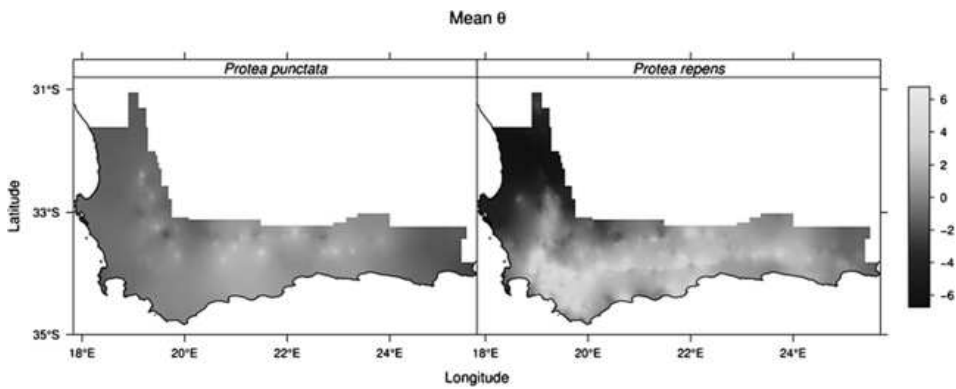}

\caption{Posterior mean spatial effects ($\theta$) for \textit{Protea
punctata} (PRPUNC) and \textit{Protea repens} (PRREPE). These effects
offer local adjustment to potential abundance. Cells with values
greater than zero represent regions with larger than expected
populations, conditional on the other covariates. }
\label{fig6}
\end{figure}

\begin{figure}

\includegraphics{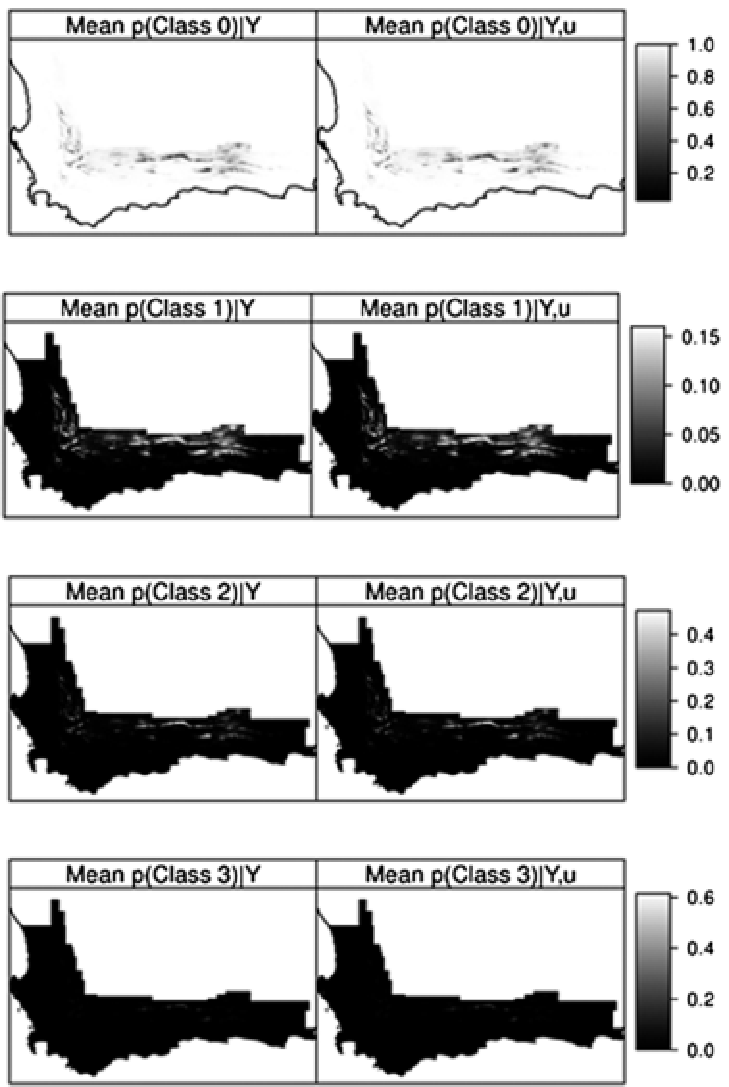}

\caption{Abundance category probability maps for \textit{Protea
punctata} (PRPUNC) for untransformed (left) and transformed (right)
situations. Values are cellwise posterior mean probabilities for the
abundance classes. Class $0$ means the probability the species is absent,
while classes $1$--$3$ indicate estimated abundance from $1$--$10$, $11$--$100$,
$100{+}$ individuals, respectively.}
\label{fig7}
\end{figure}

\begin{figure}

\includegraphics{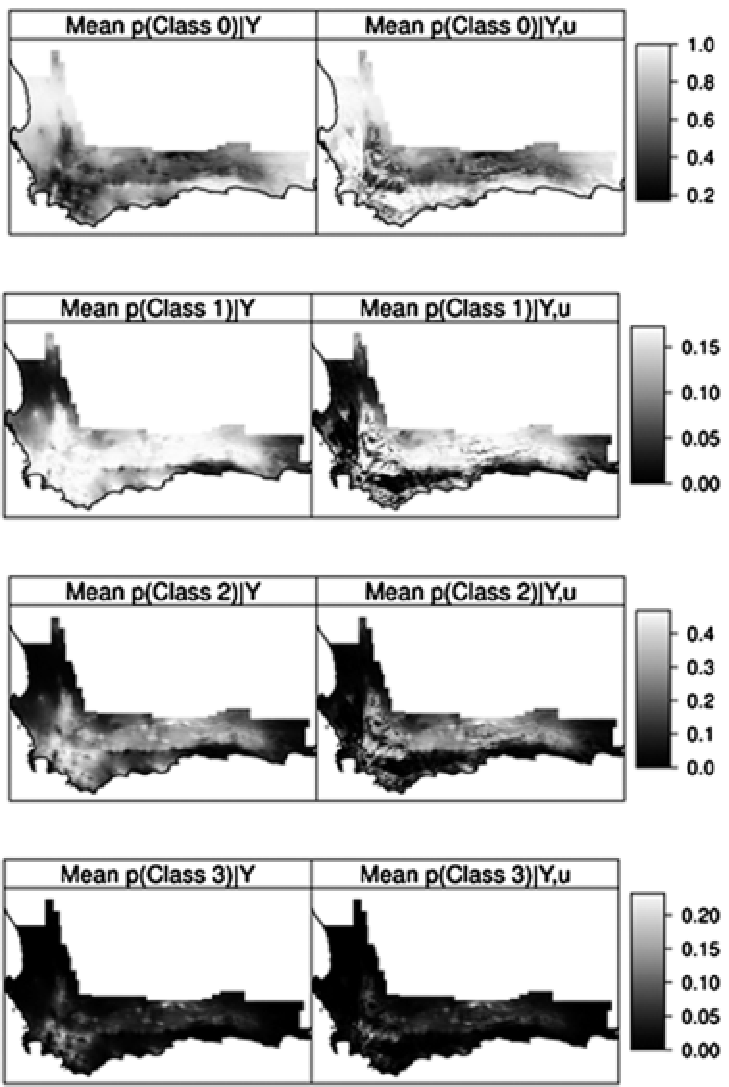}

\caption{Abundance category probability maps for \textit{Protea repens}
(PRREPE) for untransformed (left) and transformed (right) situations.
Values are cellwise posterior mean probabilities for the abundance
classes. Class $0$ means the probability the species is absent, while
classes $1$--$3$ indicate estimated abundance from $1$--$10$, $11$--$100$, $1008$
individuals, respectively.}
\label{fig8}
\end{figure}

\begin{figure}

\includegraphics{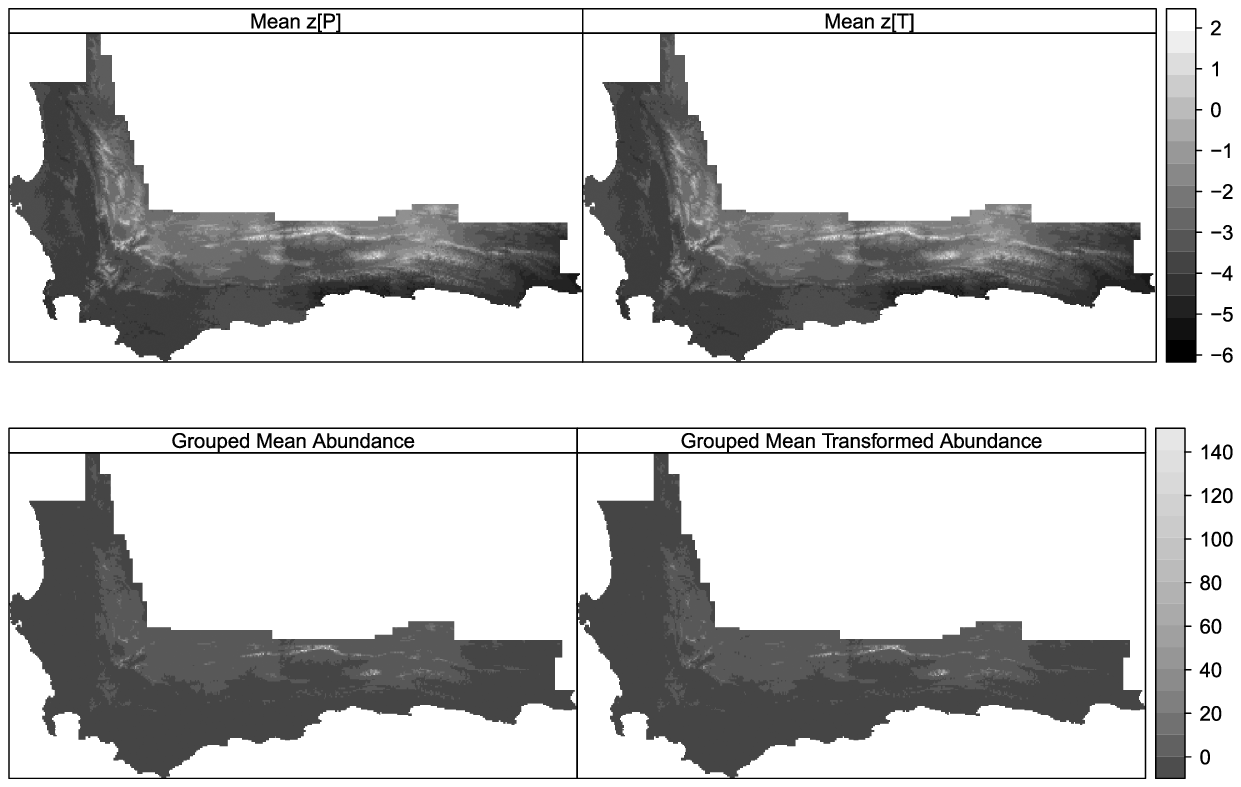}

\caption{Mean posterior abundance summaries for \textit{Protea
punctata} (PRPUNC). On the latent $z$-scale, ``Mean $z[P]$'' refers to
the potential abundance, while ``Mean $z[T]$'' refers to the potential
abundance corrected for habitat transformation. The \textit{Grouped Mean
Abundance} rescales the Mean $z[P]$ surface to the expected potential
size of a population in a grid cell (using the observed abundance
classes: absent, $1$--$10$, $10$--$100$, $100{+}$). The \textit{Group Mean Transformed
Abundance} shows the expected size of a population after correcting for
habitat transformation.}
\label{fig9}
\end{figure}

\begin{figure}

\includegraphics{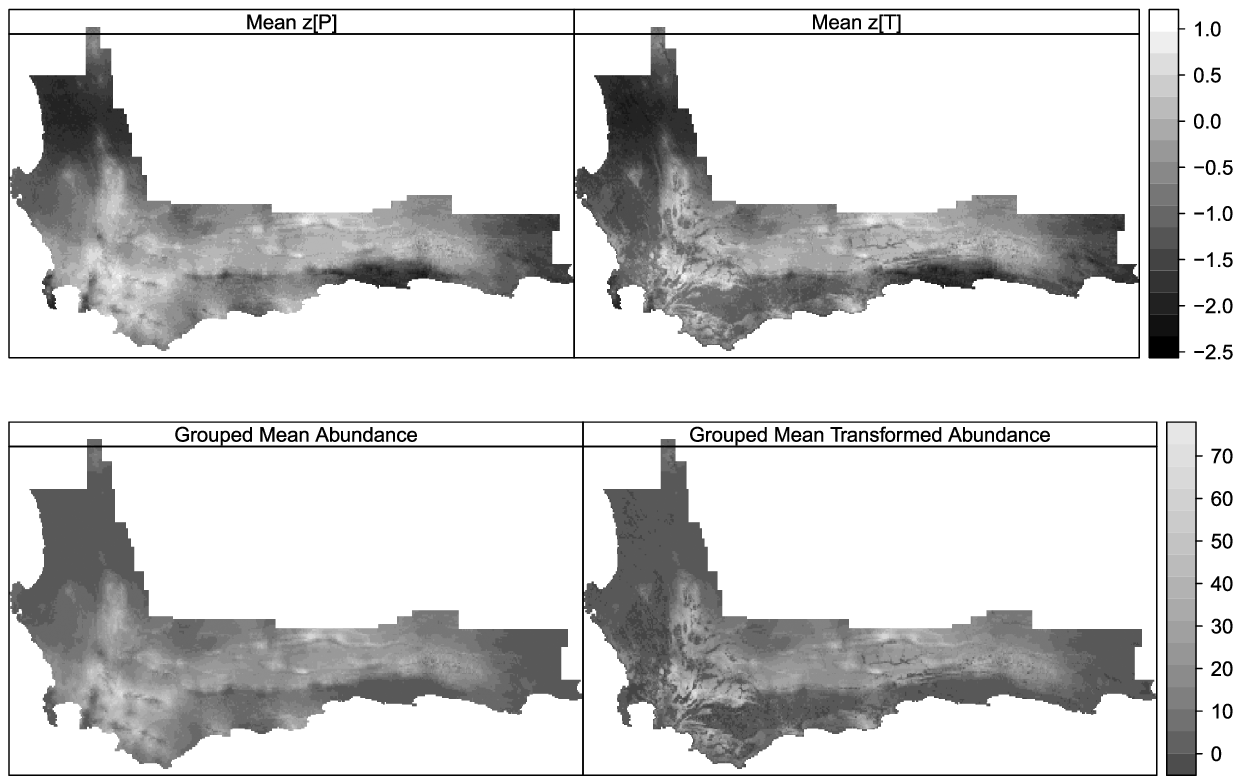}

\caption{Mean posterior abundance summaries for \textit{Protea repens}
(PRREPE). On the latent $z$-scale, ``Mean $z[P]$'' refers to the
potential abundance, while ``Mean $z[T]$'' refers to the potential
abundance corrected for habitat transformation. The \textit{Grouped Mean
Abundance} rescales the Mean $z[P]$ surface to the expected potential
size of a population in a grid cell (using the observed abundance
classes: absent, $1$--$10$, $10$--$100$, $100{+}$). The \textit{Group Mean Transformed
Abundance} shows the expected size of a population after correcting for
habitat transformation.}\label{pic1}
\label{fig10}
\end{figure}

 The mean posterior spatial effects are shown in Figure~\ref{fig6}. Note that the spatial effects for the two species have quite
different patterns, with \textit{Protea repens} having a region of low
values in the northeast and larger values elsewhere, while Protea
punctata is more even across the landscape, but with lower values
toward the edges of the \textit{CFR}. These surfaces capture the spatial
variability in abundance that is not explained by the other covariates
within the model. This suggests that the covariates predict higher
abundance of \textit{P. repens} in the northwest than what was
observed, perhaps indicating some unobserved limiting factor (such as
unsuitable soils, more extreme seasonality in rainfall, or dispersal
limitations). Similarly for \textit{P. punctata}, the covariates may
over-predict abundances at the edges of the CFR where many
environmental factors change as one transitions to other biome types.

 Figures~\ref{fig7} and~\ref{fig8} show the mean posterior
abundance category probabilities (potential and transformed)
for \textit{P. punctata} (Figure~\ref{fig7}) and \textit{P. repens} (Figure~\ref{fig8}). Comparing these plots among rows contrasts the probabilities
associated with each abundance class for the species, while comparing
between columns shows the effects of landscape transformations on
abundance class probabilities. Both species show higher predicted
abundances coinciding with mountainous areas of the \textit{CFR}. This is
where the fynbos biome dominates the landscape and where proteas are
characteristically the dominant, indicator species [\citet{Rebelo06b}].
Note that \textit{P. punctata}, a less common species, is only slightly
affected by landscape transformation, while there are dramatic
differences for \textit{P. repens} (Figures~\ref{fig9} and~\ref{fig10}). This is because \textit{P. punctata} is mostly limited to
dry, rocky, or shale slopes [\citet{Rebelo01}] which are less suitable
for agriculture or development and thus mostly untransformed. \textit{P. repens}, on the other hand, is much more ubiquitous across the
region and can frequently occur in lowland areas that have been largely
transformed by human activities [\citet{Rebelo01}, \citet{Rebelo06b}].

 It is also useful to summarize these data through mean
potential abundance and mean transformed abundance (see Section~\ref{subsec43}) as in Figures~\ref{fig9} and~\ref{fig10}. These figures
allow inspection of the underlying latent surfaces that are of interest
to ecologists as a continuous relative representation of species
abundances. However, the latent ``$z$'' scales may be difficult to
interpret ecologically and, thus, estimated potential and transformed
abundance (using the grouped mean) are also shown. These represent the
expected abundance (with respect to the $p$'s or $r$'s) of a species at a
randomly selected sample location in that grid. The associated display
makes it easy to visualize the effects of habitat transformation on
protea populations. \textit{P.~punctata} shows almost no effects of
landscape transformation, while large differences are apparent for
\textit{P. repens.} Note the large transformed regions in the south and
west where the expected abundance of plants has dropped from more than
50 to near zero. It is also apparent that, across the landscape, \textit{P. punctata} tends to have a higher expected mean abundance at any
given sample point than does \textit{P. repens} (Figures~\ref{fig9} and
\ref{fig10}).

\section{Discussion and future work}\label{sec6}

 Building on previous efforts that have addressed the
presence/absence of species, we have presented a modeling framework for
learning about potential patterns for species abundance not degraded by
land transformation and potential measurement error. The model was
built using a hierarchical latent abundance specification,
incorporating spatial structure to capture anticipated association
between adjacent locations. Along with potential pattern, we also have
an estimate of transformed abundance pattern. Comparison of these two
patterns is helpful for understanding the effect of land transformation
on species presence and abundance and, in particular, for disentangling
these effects from those of other environmental factors. This may
facilitate designing strategies for species conservation as well as
understanding the overall effects of climate change.

 This work has applications in biogeography and in
conservation biology [\citet{Pearce01}, \citet{Gaston03}].
We can now develop predictive maps of ``high quality'' habitat sites
within a species range, based on high predicted abundances. This will
help identify prime locations for effective conservation efforts. We
can also estimate the impact of habitat transformation on the size of
the population using the information from Figures~\ref{fig8} and~\ref{fig10}, and thus identify threats to conservation. Predictive
abundance maps will also be useful to explore patterns in biodiversity
and species abundances. Do species abundances tend to peak in the
middle of the species' range [\citet{Gaston03}]? Do areas of high
biodiversity tend to have lower species abundances? Are there areas
that are rich in both abundance and biodiversity (perhaps identifying
ideal regions for conservation efforts)?

 There are several natural extensions. One is to study the
temporal change in abundance. With abundance data collected over time
as well as associated environmental factors such as rainfall and
temperature, dynamic modeling of species abundance with changing
environmental factors may give a clearer picture of how a species is
responding to climate change. Indeed, when connected to future climate
scenarios, we may attempt to forecast prospective species abundance.
Similarly, if the transformation data is also time varying, we could
illuminate the effect of land transformation in greater detail.

 The current model uses transformation percentage $(1-u)$ in a
deterministic way (transformation having a binary effect on potential
abundance). In other cases (e.g., to study abundance pattern of
animals) it may be reasonable to treat transformation as another
covariate influencing species habitat. Also, it may be imagined that
the relationship between potential abundance and environmental
variables is not linear as specified in equation~(\ref{eq331}), for example, environmental
variables may affect larger abundance classes differently from smaller
abundance classes; piecewise linear specification, introducing
different regression coefficients over the different abundance classes,
could be explored.

 Another possible extension lies in joint modeling of two or
more species. One may wish to learn whether two plant varieties are
sympatric or allopatric and whether or not there is evidence for
competitive interactions or facilitation. Such modeling can be done by
extending our model to have multiple $(z_{P,k},z_{T,k},z_{O,k})$
surfaces, where $k$ is the species indicator. Dependence can be
introduced across $z_{P,k}$ surfaces by modeling $\theta^{(k)}$ using
an MCAR [\citet{Gelfand03}, \citet{Jin05}]. Fitting such models will be very
challenging if there are many grid cells.

 Instead of taking an areal level approach, if covariate
information is available at point level (where sampling sites are
viewed as ``points'' within the large region,~$D$), one may consider a
point-level model. This amounts to replacing the \textit{CAR} model with a
Gaussian process prior for the spatial effects. With many sampling
sites, we will need to use appropriate approximation techniques [\citet{Banerjee08}].

\begin{appendix}

\section*{Appendix}\label{app}
\subsection{Proof of $E(z_T)$ finite}\label{A1}
$E(z_T) = E(E(z_T|z_P)) = E(uz_P + (1-u)\times\break c(z_P)) = E(z_P - (1-u)\frac{\phi(z_P)}{1-\Phi(z_P)})$.
Assuming\vspace*{-2pt} $z_P\sim N(\mu,1)$, it is enough
to show $\int_{-\infty}^{\infty}\frac{\phi(x)}{1-\Phi(x)}\phi(x-\mu)\,dx < \infty$.

 Consider the quantity $x^2\frac{\phi(x)}{1-\Phi(x)}\phi(x-\mu
)$, if $x\rightarrow-\infty$, it goes to $0$. When $x \rightarrow
\infty$, we have
\begin{eqnarray*}
&&\hspace*{-4pt} \lim_{x\rightarrow\infty}x^2\frac{\phi(x)}{1-\Phi(x)}\phi(x-\mu)
\\
&&\hspace*{-4pt}\qquad\stackrel{L'ptal}{=} \lim_{x\rightarrow\infty}\frac{2x\phi(x)\phi
(x-\mu) - x^3\phi(x)\phi(x-\mu) - x^2(x-\mu)\phi(x)\phi(x-\mu)}{-\phi(x)}\\
&&\hspace*{4pt}\qquad= 0.
\end{eqnarray*}
 So $\lim_{|x|\rightarrow\infty}x^2\frac{\phi(x)}{1-\Phi(x)}\phi(x-\mu)=0$, thus, we can get $B_1<0,B_2>0$, such that $\frac
{\phi(x)}{1-\Phi(x)}\phi(x-\mu)<\frac{1}{x^2}$ for all $x \notin
(B_1,B_2)$. Hence, the result follows.

\subsection{Posterior simulation of $z$'s for a site with no presence
observed}\label{A2}
 We subdivide by considering the ways that we can generate a
$0$ realization of $y$ based on Equation (\ref{eq333}) [one may also
use Equation (\ref{eq331}) to do this]:
\begin{enumerate}[(iii)]
\item[(i)] The area is untransformed, the species was potentially
there, but missed during data collection or it was absent at that time
instance; the event is $1_{z_{P}\geq\alpha_0,z_{O}\leq\alpha_0}$ with
prior probability $\pi_1 = uP(z_{P}\geq\alpha_0,z_{O}\leq\alpha_0)$.
\item[(ii)] Potentially the species was absent there; the event is
$1_{z_{P}\leq\alpha_0}$ with prior probability $\pi_2 = P(z_{P}\leq
\alpha_0)$.
\item[(iii)] The species was potentially there $1_{z_{P}\geq\alpha
_0}$, but the area was transformed; the event has prior probability $\pi
_3 = (1-u)P(z_{P}\geq\alpha_0).$
\end{enumerate}

 These three events are exhaustive and mutually exclusive for
the event $(y=0)$. Thus, $f(z_P,z_O|y=0,\Theta)$ is a $3$-component
mixture. To draw a ($z_P,z_O$) pair from this distribution amounts to
first choosing a component and then drawing a pair ($z_P,z_O$) from
that component distribution. By Bayes' rule, conditional on observed
$(y=0)$, these three cases can happen with posterior probability $\pi
_i/ \sum_{l=1}^{3} \pi_l, i=1,2,3$. So we use a
multinomial to select which of these events took place. Before going
into case by case details, it is worth mentioning that in all these
cases the sampling from the joint density of chosen mixture component
was done via the marginal
$f(z_P|\cdot \cdot)$ followed by
$f(z_O|z_P,\cdot\hspace*{-1pt}\ \cdot)$. The advantage of this scheme is that we don't
need to draw from the latter because $z_O$'s corresponding to $y=0$ are
not involved in posterior full conditionals of any other parameters in
the model (as $\alpha_0 = 0$, fixed). If the second case is selected,
then $f(z_{O},z_{P}|\cdot,\cdot)\propto [ u\delta_{z_{P}}+(1-u)\delta_{c(z_P)} ]1_{z_{O} \leq0}1_{z_{P} \leq0}\phi
(z_P)$ and thus marginalizing over $z_{O}$, we get
$f(z_P|\cdot \cdot)\propto\phi(z_P)1_{z_{P} \leq0}$ which is a truncated Gaussian on
$\mathbb{R^{-}}$. Similarly under case (iii), we need to simulate
$z_P$ from $\phi(z_P)1_{z_{P} \geq0}$, a Gaussian truncated on $\mathbb{R^{+}}$. In case (i), $f(z_{O},z_{P}|\cdot\cdot)\propto\phi
(z_{O};z_{P},1)1_{z_{P} \geq0}1_{z_{O} \leq0}\phi(z_P)$, so marginalizing over $z_{O}$ we get
$f(z_P|\cdot \cdot)\propto\phi
(z_P)(1-\Phi(z_P))1_{z_{P} \geq0}$. An efficient way to draw from this
density is to propose a $z_P$ from a truncated normal on $\mathbb{R^{+}}$ and do a Metropolis--Hastings update with an independent
proposal, using the quantity $(1-\Phi(\cdot))$. However, all sampling
distributions are summarized in~Appendix~\ref{A3} below.

\subsection{Posterior full conditionals needed for Gibbs sampling}\label{A3}
\begin{itemize}
\item If $y_{ij}>0$, draw $z_{O,ij}\sim N(z_{P,ij},1)1_{(\alpha
_{y_{ij}-1},\alpha_{y_{ij}})}$. Draw $z_{P,ij}\sim N(\frac{v_i^T\beta+
\theta_i}{2} + \frac{z_{O,ij}}{2},\frac{1}{2})$.
\item If $y_{ij}=0$, compute $p_{ij} = (u\Phi_2( [ 0,\infty
]\times [- \infty,0  ];\mu_{ij},\Sigma_0),1-\Phi(v_i^T\beta+
\theta),(1-u)\Phi(v_i^T\beta+ \theta_i))$, where $\mu_{ij}=
(v_i^T\beta+ \theta_i,v_i^T\beta+ \theta_i)$ and $\Sigma_0 =
 \left({ {1 \atop 1}\enskip{1 \atop 2}}\right) $
 are the location\vspace*{-2pt} and dispersion parameters for bivariate
normal joint prior distribution of $(z_{O,ij},z_{P,ij})$. Draw $d_{ij}
\stackrel{\mathrm{i.i.d.}}{\sim} \operatorname{mult}(p_{ij})$. If $d_{ij}=1$, propose
$z_{P,ij}^{\mathit{propose}}\sim N(v_i^T\beta+ \theta_i,1)1_{(0,\infty)}$ and
do a Metropolis--Hastings sampler using $(1-\Phi(\cdot))$. Else if
$d_{ij}=2$, draw $z_{P,ij}\sim N(v_i^T\beta+ \theta_i,1)1_{(- \infty
,0)}$, else draw $z_{P,ij}\sim N(v_i^T\beta+ \theta_i,1)1_{(0,\infty)}$.
\item Draw $\alpha_h = \operatorname{unif}(\max_{ij : y_{ij} = h} z_{O,ij},
\min_{ij:y_{ij}=h+1} z_{O,ij})$, $h=1,2$.\vspace*{1.5pt}
\item Draw $\beta\sim$ $N(\mu_{\beta},\Sigma_{\beta})\prod_{i,j}
N(z_{P,ij}; v_i,\beta, \theta_i)$.
\item Draw\vspace*{-2pt} $\theta_i \sim$ $N(z_{P,ij}; v_i,\beta,\theta_i)N(\frac{\sum
_{j}w_{ij}\theta_j}{w_{i+}},\frac{\tau_{0}^2}{w_{i+}})$ for
$i=1,2,\ldots,m$. Draw $\theta_i \sim$ $N(\frac{\sum_{j}w_{ij}\theta
_j}{w_{i+}},\frac{\tau_{0}^2}{w_{i+}})$ for $i=m+1,2,\ldots,I$.
\end{itemize}
\end{appendix}

\section*{Acknowledgments}

The authors thank Guy Midgley and Anthony Rebelo for useful discussions.

\printaddresses

\end{document}